# Designing a Cost-Time-Quality-Efficient Grinding Process Using MODM Methods

Meysam Mahjoob and Pouneh Abbasian

***Abstract* — In this paper, a multi-objective mathematical model is developed to optimize the grinding parameters Such as grinding time, cost, and related surface quality metrics such as workpiece speed, depth of cut and wheel speed. The mathematical model consists of three conflicting objective functions subject to wheel wear and production rate constraints. Exact methods cannot solve the NLP model in few seconds, therefore using Meta-heuristic algorithms that provide near-optimal solutions is not suitable. Considering this, five multi-objective decision-making (MODM) have been used to solve the multi-objective mathematical model using general algebraic modelling system (GAMS) software to achieve the optimal parameters of the grinding process. The MODM methods provide different effective solutions where the decision-maker (DM) can choose each solution in different situations. Different criteria have been considered to evaluate the performance of the five MODM methods. Also, a technique for order of preference by similarity to ideal solution (TOPSIS) has been used to obtain the priority of each method and determine which MODM method performs better considering all criteria simultaneously. The results indicated that the weighted sum method (WSM) and goal programming method (GP) are the best MODM methods, as both of them provide competitive solutions. In addition, these methods obtained solutions that have minimum grinding time, cost, and surface roughness among other MODM methods.**

## I. INTRODUCTION

During the past decade, many researchers have focused on optimizing the grinding process. Production costs, machining time, and surface quality of machined products can be improved by achieving optimal grinding process parameters [1].

Baskar *et al.* [2] proposed an ant colony-based optimization approach to optimize the grinding parameters using a multi-objective model with a weighted method under thermal damage, wheel wear parameter, surface finish, and tool stiffness constraints. They compared the results with Quadratic programming (QP) and Genetic Algorithm (GA) presented in previous researches. They showed that the ant colony-based optimization method performs better in solving the grinding process optimization problem. Saravanan *et al.* [3] proposed a new GA to solve the weighted objective function of the grinding optimization problem. The results declared that their approach is a robust and easy method compared to the previous studies. More researches in the optimization of the grinding process include [4]-[19]. Diverse optimization methods are suggested considering the effect of the grinding parameters such as wheel speed, workpiece speed, depth of dressing, lead of dressing on the manufactured products. Gholami and Azizi [20] presented a non-dominated sorting genetic algorithm (NSGA II) to obtain the optimal values of workpiece speed, wheel speed, and depth of cut in the grinding process. They presented different Pareto solutions for the multi-objective optimization problem selected by the decision-maker (DM) under different scenarios. Multi-choice goal programming (MCGP) is another concept that provides a range of ideal solutions for each objective function [21], thereby it is more flexible than goal programming (GP) in situations when DM underestimates the initial ideal solution set for the model. lack of available recourses and information could be reasons for changes in DM’s preferences in different situations and times. Using multiple utility functions defined for each objective function, [22], [23] were able to solve multi-objective decision making problems using the Bayesian theory.

The majority of previous studies combined the objective functions to construct a single weighted objective function in order to optimize the grinding parameters. This conversion may lead to significant deviations in obtaining the optimal value of the decision variables and the solution's quality. Also, the solutions' quality strongly depends on the weight assigned to each objective function, where finding suitable weights for each objective function is another complex decision. Moreover, exact methods can solve the NLP model of the grinding process in a few seconds. Therefore, using Meta-heuristic algorithms that provide near-optimal solutions is not suitable. There are many other solution methods that handle multi-objective optimization problems, such as multi-objective decision making (MODM) methods [22].

As we mentioned earlier, in the case of not-so-complex problems, the exact methods would solve problems in a few seconds. Therefore, five MODM methods have been used to solve the multi-objective mathematical model of the problem using the general algebraic modeling system (GAMS) software which provides exact solutions for optimization problems. The solution obtained by each method is an effective solution to the optimization problem and the DM can choose each MODM method in different situations. In addition, MODM methods can provide better solutions than

Meysam Mahjoob, Department of Industrial and Systems Engineering, Fouman Faculty of Engineering, College of Engineering, University of Tehran, Fouman, Iran.

(e-mail: mahjoob_m@ut.ac.ir)

Pouneh Abbasian, Department of Multidisciplinary Engineering, College of Engineering, Texas A&M University, USA.

(e-mail: p.abbasian@tamu.edu)

meta-heuristic algorithms such as NSGA-II, MOPSO and etc. At the end we compared the performance of the MODM methods using different criteria including objective functions value and CPU-Time. Technique for order of Preference by Similarity to Ideal Solution (TOPSIS) method has been used to determine the best MODM method in solving the multi-objective mathematical model of the grinding process.

## II. Mathematical Model

We used the multi-objective mathematical model of the grinding parameters proposed by [20]. The mathematical model of the problem includes three objective functions, a non-linear constraint and upper and lower bounds for the decision variables. In this research, the following notation has been used:

*$M_c$: Cost per hour of labor and administration (\$/h)*
*$p$: Number of workpieces loaded on the table*
*$L_w$: Length of workpiece (mm)*
*$L_e$: Empty length of grinding (mm)*
*$b_w$: Width of workpiece (mm)*
*$b_e$: Empty width of grinding (mm)*
*$f_b$: Cross feed rate (mm/pass)*
*$a_w$: Total thickness of cut (mm)*
*$a_p$: Down feed of grinding (mm/pass)*
*$S_p$: Number of sparks out grinding (pass)*
*$d_e$: Diameter of wheel (mm)*
*$b_s$: Width of wheel (mm)*
*G: Grinding ratio*
*$S_d$: Distance of wheel idling (mm)*
*$V_r$: Speed of wheel idling (mm/min)*
*$t_i$: Time of loading and unloading workpiece (min)*
*$t_{ch}$: Time of adjusting machine tool (min)*
*$N_d$: Total number of pieces to be grouped during the life of dressing*
*$N_t$: Batch size of workpiece*
*$N_{td}$: Total number of workpieces to be grouped during the life of dressing*
*$C_d$: Cost of dressing (\$)*
*$C_s$: Cost of wheel per $mm^3$ (\$/$mm^3$)*
*$C_T$: Production cost (\$)*
*$R_a$: Surface roughness ($\mu m$)*
*Doc: Depth of dressing (mm)*
*L: Lead of dressing (mm/rev)*
*$WRP$: Workpiece removal parameters ($mm^3$/min N)*
*$WWP$: Wheel wear parameter (mm3/min N)*
*$T$: Total grinding time*
*$N_p$: Number of passes*
*$t_m$: Time of machining (min)*
*$t_a$: Time of workpiece approach (min)*
*$V_s$: Wheel speed (m/min)*
*$t_e$: Extra workpiece path time (min)*
*$V_w$: Work piece speed (m/min)*

As in [20], the mathematical model of the problem can be represented as:

$$Min\ R_a = 4.456 V_w^{0.229} . a_w^{-1.649} . V_s^{-0.964} \tag{1}$$

$$MinT = \frac{N_p \times L_w}{V_w} + t_{ch} + t_i + \frac{L_e}{V_w} + \frac{L_e}{V_w} \tag{2}$$

$$Min\ C_T = \frac{M_c}{60p}\left(\frac{L_w+L_e}{1000V_w}\right)\left(\frac{b_w+b_e}{f_b}\right)\left(\frac{a_w}{a_p} + S_p + \frac{a_w b_w L_w}{\pi D_e b_s a_p G}\right) + \frac{M_c}{60p}\left(\frac{S_d}{V_r} + t_1\right) + \frac{M_c}{60}\frac{t_{ch}}{N_t} + \frac{M_c}{60p}\frac{1}{N_d}\frac{\pi D_e b_s}{1000LV_s} + C_s\left(\frac{a_w b_w L_w}{pG} + \frac{\pi Doc b_s D_e}{pN_d}\right) + \left(\frac{C_d}{pN_{td}}\right) \tag{3}$$

Subject to:

$$WRP = 94.4 \frac{\left(\frac{2Doc}{3L}+1\right) L^{\frac{11}{19}} \left(\frac{V_w}{V_s}\right)^{\frac{3}{19}} V_s}{D_e^{\frac{43}{304}} VOL^{0.47} d_g^{\frac{5}{38}} R_c^{\frac{27}{19}}} \tag{4}$$

$$WWP = \frac{K_a a_p d_g^{\frac{5}{38}} R_c^{\frac{27}{19}}}{D_e^{\left(\frac{1.2}{VOL} - \frac{43}{304}\right)} VOL^{0.38}} \frac{\left(1+\frac{Doc}{L}\right) L^{\frac{27}{19}} \left(\frac{V_s}{V_w}\right)^{\frac{3}{19}} V_w}{\left(1+\frac{2Doc}{3L}\right)} \tag{5}$$

$$G \leq \frac{WRP}{WWP} \tag{6}$$

$$1000 \leq V_s \leq 3000 \tag{7}$$

$$10 \leq V_w \leq 50 \tag{8}$$

$$0.04 \leq a_w \leq 0.12 \tag{9}$$

The aims of (1)-(3) are to minimize the production costs, grinding time and surface roughness simultaneously. Inequality (6) determines the wheel wear constraint and (7)-(9) indicates the upper and lower bounds of each decision variables.

## III. Solution Methods

The mathematical model developed in the previous section is a constraint bi-objective mixed integer linear programming (MILP) model. The optimal solution of the developed bi-objective model is an ideal solution that minimizes both objective functions simultaneously. Since the objective functions are in conflict and so there is no such optimal solution, thus we need to make a compromise solution between the objectives [24]-[26]. In these cases, the multi-objective solution methods should be utilized to solve the model. In this paper five MODM methods presented by [27] utilized to solve the multi objective optimization model of the grinding process. As in [28]-[30], five MODM methods are defined as follows:

### A. Individual Optimization Method

This method considers each objective function separately, solves the optimization problem and obtains the optimal solution. This method is based on this concept that the optimal solution of each objective function is an effective solution for the multi-objective optimization problem.

### B. Lp-Metric Method

This method is based on the concept of minimizing the digression between objective functions and their ideal solution obtained by the individual optimization method.

Equation (10) describes the Lp-Metric method. Minimization type objective functions must be converted to maximization type.

$$Min\ D = \left( \sum_{i=1}^{n} \left( \frac{f_i^* - f_i}{f_i^*} \right)^r \right)^{\frac{1}{r}} \quad (10)$$

### C. Weighted Sum Method (WSM)

In WSM method, a positive weight is assigned to each objective function. The assigned weights to objective functions must satisfy the $\sum_{i=1}^{n} w_i\, f_i = 1$ constraint. The goal is to minimize the combined objective function which is weighted sum of the objective functions as following:

$$Max\ U\ (f_1, f_2, \ldots, f_n) = \sum_{i=1}^{n} w_i\, f_i \quad (11)$$

### D. Max-Min Method

The purpose of Max-Min method is to maximize the minimum values of objective functions divided to their ideal solutions. Equation (12) indicates the mathematical model of the method.

$$Max\ \left( Min \left( \frac{f_1}{f_1^*}, \frac{f_2}{f_2^*}, \ldots, \frac{f_n}{f_n^*} \right) \right) \quad (12)$$

### E. Goal Attainment Method

The Goal attainment method aims to find solutions for each objective function which minimizes a weighted deviation of objective function values with their related ideal solution. The assigned weights to deviations in objective functions must satisfy the $\sum_i w_i = 1$ constraint. The mathematical model of the problem is as follows:

$$\begin{aligned} & Min\ \ Z \\ & s.t.: \\ & f_i + w_i Z \geq f_i^*\ ; \forall\, i \end{aligned} \quad (13)$$

### F. Goal Programming Method (GP)

In GP method the aim is to find a solution which minimizes the positive or negative deviations between objective functions and their relevant ideal solutions. Equation (14) defines the mathematical model of the Goal Programming method.

$$\begin{aligned} & Min\ \ \sum_{i=1}^{n} a_i\, g_i\, (d_i^+, d_i^-) \\ & s.t. \\ & f_i - d_i^+ + d_i^- = f_i^*\ ; \forall\, i \\ & d_i^+ \geq 0\ ,\ d_i^- \geq 0\ ; \forall\, i \end{aligned} \quad (14)$$

## IV. Experimental Example

Gholami and Azizi [20] used nine billets of 1.2080 steel with $30 \times 20 \times 10$ mm dimensions to perform the grinding process. The material of the abrasive grinding wheel has been selected from aluminum oxide. In order to dress the grinding wheel a single point diamond dresser has been utilized. To optimize the grinding process the values of the input parameters have been selected as [20] in Table I.

Five MODM methods presented above to optimize the grinding parameters have been applied to achieve the best finish surface, minimum grinding cost and time using GAMS software. For this purpose, a computer with i7 CPU and 8GB of ram has been utilized. Different criteria have been considered to evaluate the performance of the five MODM methods such as objective function values and CPU-Time [31].

### A. Objective Functions Value

The three objective functions value have been considered as three different criteria to compare the MODM methods in term of ability to achieve the best optimal solution [32], [33].

### B. CPU-Time

CPU-Time criterion is another important factor to compare the MODM methods in term of time needed to solve the multi-objective optimization problem [34]-[36].

The result of solving the optimization problem using five MODM methods is presented in Table II.

TABLE II: RESULTS OF MODM METHODS

| Method | $V_w$ | $V_s$ | $a_w$ | $R_a$ | T | $C_T$ | CPU-Time |
|---|---|---|---|---|---|---|---|
| Individual optimization method | --- | -- | -- | 0.111 | 25 | 5.336 | -- |
| Lp-Metric | 31.60 | 3000 | 0.12 | 0.144 | 26.7 | 5.656 | 0.100 |
| Max-Min | 10 | 1000 | 0.109 | 0.375 | 37 | 7.149 | 0.270 |
| Goal attainment | 10 | 1000 | 0.047 | 1.508 | 37 | 6.733 | 0.110 |
| WSM | 50 | 3000 | 0.12 | 0.16 | 25 | 5.445 | 0.094 |
| Goal programming | 50 | 3000 | 0.12 | 0.16 | 25 | 5.445 | 0.098 |

As in Table II, each solution obtained by each MODM method is an effective solution for the optimization problem. Each solution can be preferred by the DM in different situations. For example, when the importance of the surface quality is higher for the decision maker, he/she will choose the Lp-Metric method which obtains a solution with the minimum $R_a$ comparing to other methods. Fig. 1, 2 and 3 show the objective function values obtained by each MODM method. Fig. 4 depicts the CPU-Time of the MODM methods.

TABLE I: VALUES OF THE PARAMETERS

| $M_c$ | p | $L_e$ | $L_w$ | $b_w$ | $b_e$ | $f_b$ | G | $N_{td}$ | $d_e$ | $V_{w\min}$ | $V_{w\max}$ |
|---|---|---|---|---|---|---|---|---|---|---|---|
| 30 | 1 | 15 | 30 | 20 | 10 | 2 | 60 | 2000 | 500 | 10 | 50 |
| $C_T^*$ | $R_c$ | $t_e$ | $V_r$ | $t_i$ | $t_{ch}$ | $N_d$ | $N_t$ | $N_p$ | $C_d$ | $a_{w\min}$ | $a_{w\max}$ |
| 3 | 58 | 0.02 | 25 | 2 | 20 | 4 | 4 | 4 | 75 | 0.04 | 0.12 |
| $t_a$ | $s_p$ | $b_s$ | Doc | L | $K_a$ | vol | $d_g$ | $S_d$ | $C_s$ | $V_{s\min}$ | $V_{s\max}$ |
| 0.02 | 3 | 50 | 0.02 | 0.02 | 0.0869 | 6.99 | 0.3 | 10 | 0.003 | 1000 | 3000 |

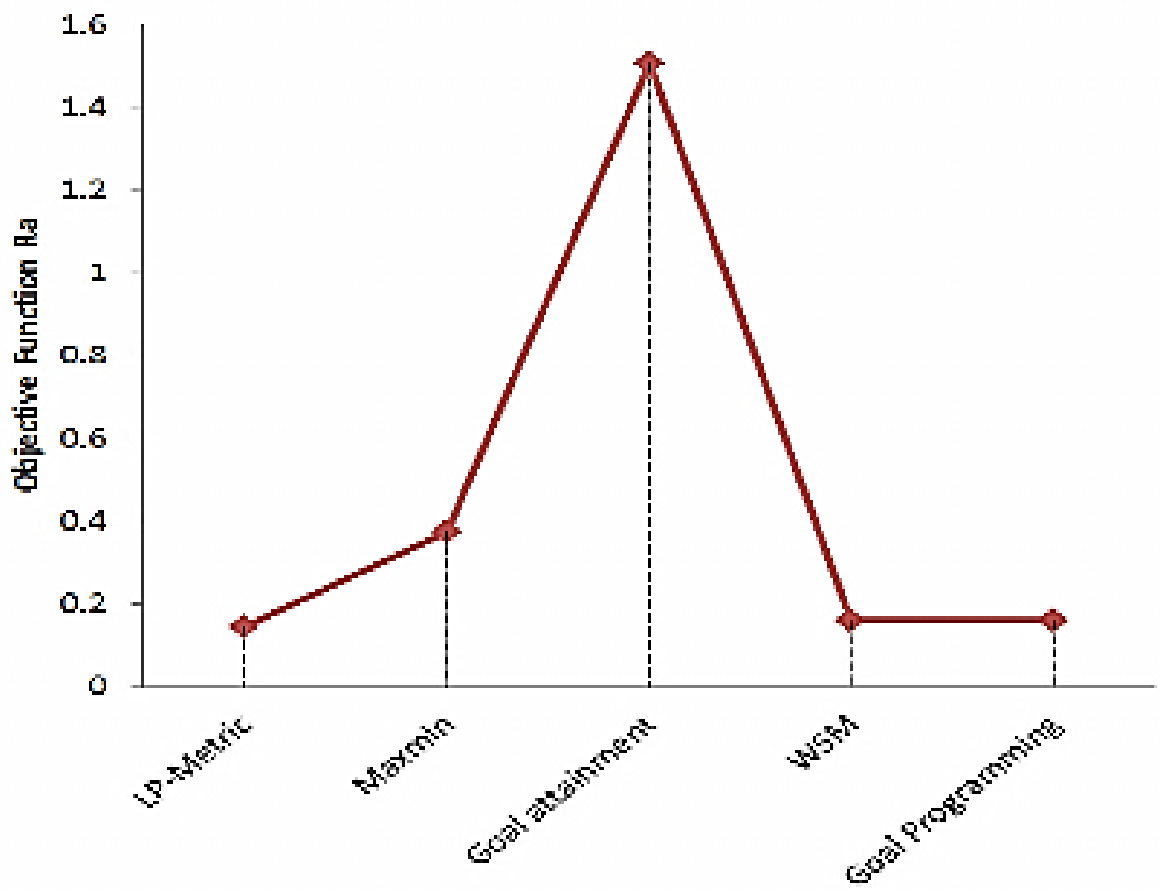

Fig. 1. Minimum surface roughness achieved by MODM methods.

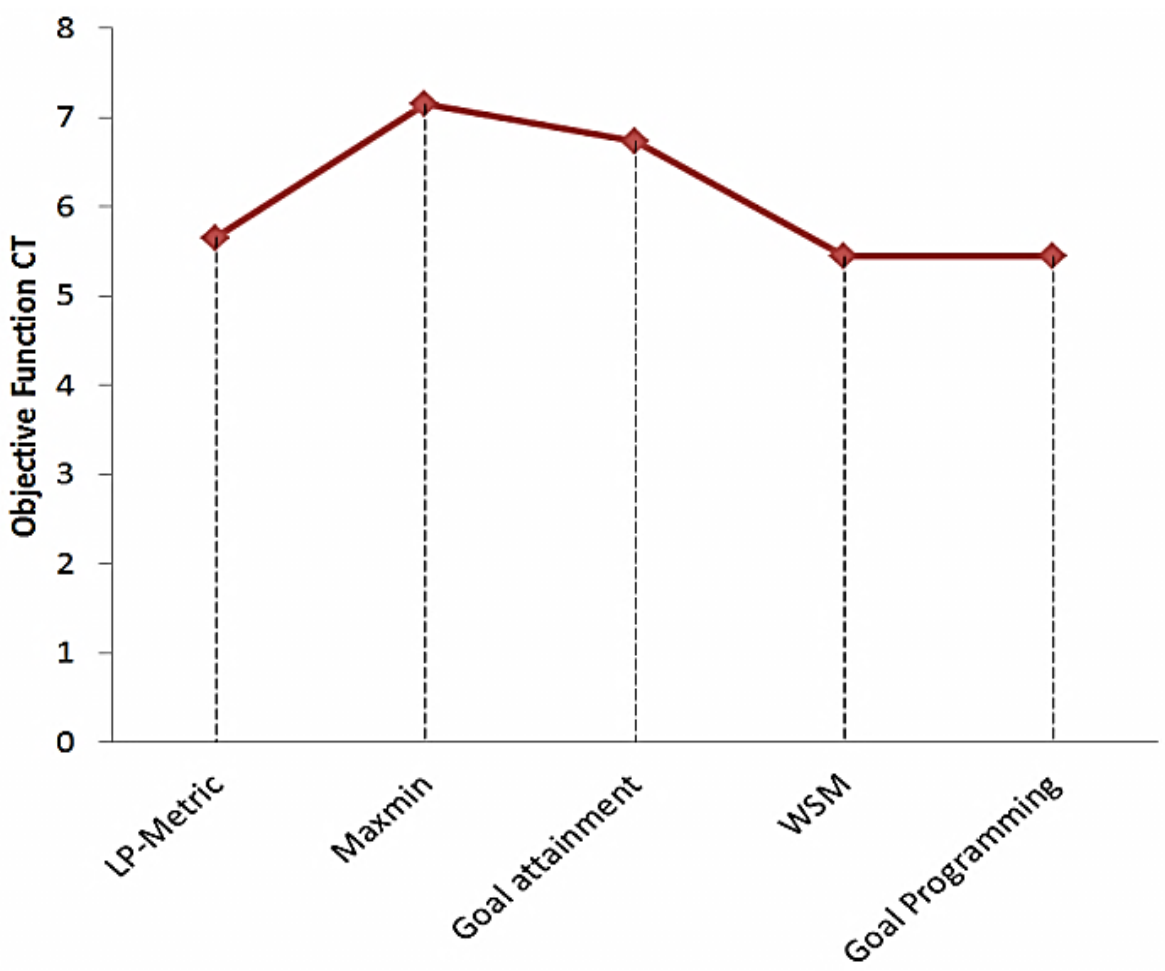

Fig. 2. Minimum production cost obtained using each MODM method.

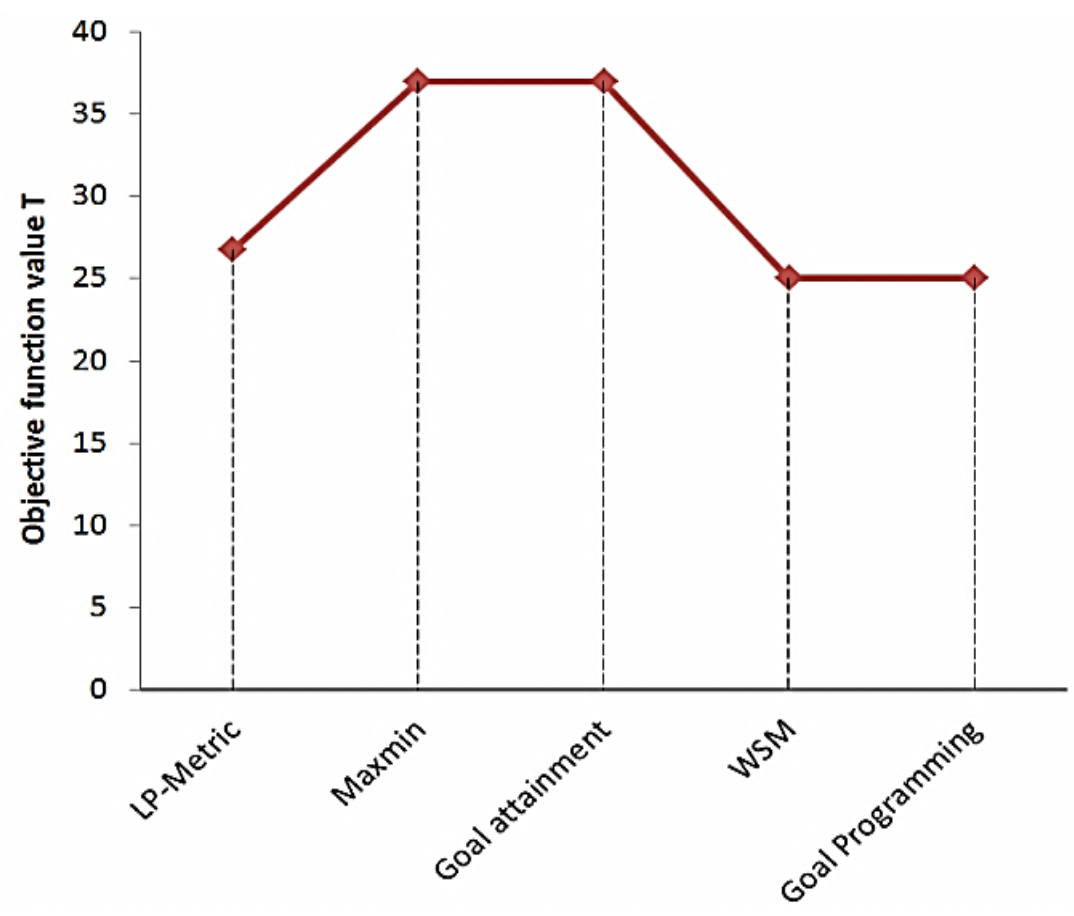

Fig. 3. Minimum total grinding time gained by MODM methods.

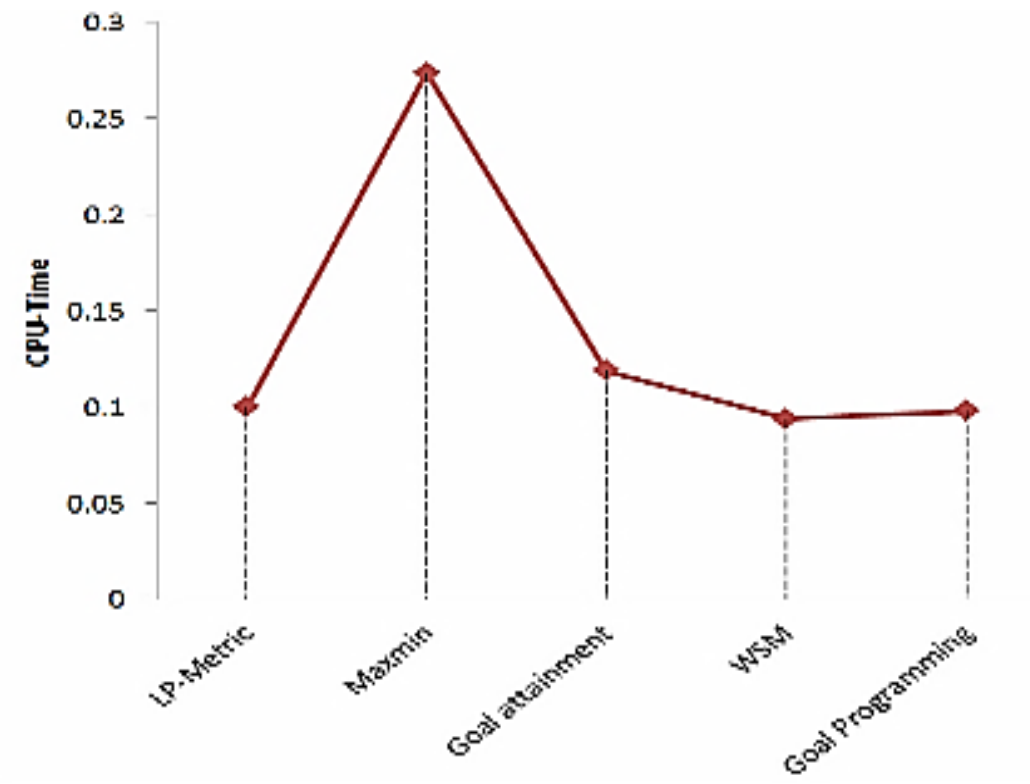

Fig. 4. CPU-Time of each MODM method.

## V. Comparing MODM Methods Using TOPSIS Method

In order to compare MODM methods, first we must build the decision matrix as presented in Table III.

TABLE III: Decision Matrix

| Method \ Criteria | $R_a$ | T | $C_T$ | CPU-Time |
|---|---|---|---|---|
| Lp-Metric | 0.144 | 26.7 | 5.656 | 0.100 |
| Max-Min | 0.375 | 37 | 7.149 | 0.270 |
| Goal attainment | 1.508 | 37 | 6.733 | 0.110 |
| WSM | 0.16 | 25 | 5.445 | 0.094 |
| Goal programming | 0.16 | 25 | 5.445 | 0.098 |

### A. *TOPSIS Method*

TOPSIS method proposed by [37]. The concept of TOPSIS method is based on selection of an alternative which has longest (shortest) distance from the negative (positive) ideal solution. TOPSIS method has been applied to determine the best MODM method in solving the multi-objective optimization problem. As objective function values are more important to us than CPU-Time, we allocated the 80% weight for objective functions criteria and 20% weight for CPU-Time criterion. The weight of each criterion is given in Table IV.

TABLE IV: Weights of the Criteria

| Criteria | $R_a$ | T | $C_T$ | CPU-Time |
|---|---|---|---|---|
| $\text{Weight}_j$ | 0.266 | 0.266 | 0.266 | 0.20 |

First, we need to normalize the decision matrix using Euclidean Norm:

$$n_{ij} = \frac{r_{ij}}{\sqrt{\sum_i {r_{ij}}^2}} \tag{15}$$

Where $r$ is the decision matrix values and $n$is normalized decision matrix using Euclidean Norm. $i$ is our chosen MODM method and$j$is the criterion index. To obtain a weighted normalized decision matrix, $\text{Weight}_j$should be multiplied by a normalized decision matrix, as shown in (16).

$$weighted\ normalized\ matrix = \left[v_{ij}\right]_{m\times n} \quad , \quad v_{ij} = \text{Weight}_j \times n_{ij} \tag{16}$$

where, $\text{Weight}_j$is the weight of each of the MODM methods. Therefore, we can determine the ideal positive solution and the ideal negative solution as follows:

$$idealsolution^{+} = \left\{max_i\, v_{ij} \quad : \quad j \in j^{+}, \quad min_i\, v_{ij} \quad : \quad j \in j^{-}\right\} \tag{17}$$

$$idealsolution^{-} = \left\{max_i\, v_{ij} \quad : \quad j \in j^{-}, \quad min_i\, v_{ij} \quad : \quad j \in j^{+}\right\} \tag{18}$$

Distance from the positive and negative ideal solutions for each MODM method have been calculated using below formulas:

$$d_i^+ = \sqrt{\sum_{j=1}^{n}(v_{ij} - idealsolution^+)} \quad (19)$$

$$d_i^- = \sqrt{\sum_{j=1}^{n}(v_{ij} - idealsolution^-)} \quad (20)$$

Equation (21) presents the Similarity ratio formula.

$$S_i^+ = \frac{d_i^-}{d_i^- + d_i^+} \quad (21)$$

The results achieved from TOPSIS method are presented in Table V.

TABLE V: TOPSIS METHOD RESULTS

| Similarity ratio | Lp-Metric | Max-Min | Goal attainment | WSM | Goal programming |
|---|---|---|---|---|---|
| $S_i^+$ | 0.9677 | 0.6037 | 0.2876 | 0.9896 | 0.9861 |

The MODM method with larger similarity ratio performs better in solving the mathematical model of the multi-objective optimization problem of the grinding process. Table VI presents five MODM methods ranked according to their similarity ratio.

TABLE VI: MODM METHODS RANKING

| Method | Rank |
|---|---|
| Lp-Metric | 3 |
| Max-Min | 4 |
| Goal attainment | 5 |
| WSM | 1 |
| Goal programming | 2 |

The results indicate that the WSM provides the best solution to the multi-objective optimization problem. Also, GP performs significantly better than other MODM methods in solving optimization problem of the grinding process.

## VI. CONCLUSION

In this paper a multi objective mathematical model have been used to optimize the grinding parameters in an experimental case study to achieve best possible grinding surface, minimum production time and cost. Combining objective functions using weighted approaches may lead to significant deviations in obtaining the optimal values of the decision variables and the quality of the solution. To avoid this, we used five different MODM methods to solve the multi objective optimization problem. Different criteria such as objective functions value and CPU-Time have been considered to compare these MODM methods. The results indicated that the solutions obtained by each MODM method is an effective solution for the multi objective model and the decision maker can choose each MODM method in different situations. TOPSIS method has been utilized to determine the best MODM method considering comparing criteria simultaneously. The results indicated that the WSM and GP methods are the best MODM methods in solving multi objective optimization problem of the grinding process.